\documentclass[twocolumn]{aastex63}
\usepackage{graphicx}	
\usepackage{amsmath}	
\usepackage{amssymb}	
\usepackage{color}
\usepackage{array}
\usepackage{lineno}
\usepackage{bookmark}
\usepackage{subfigure}

\shorttitle{GW memory from GRB afterglows with energy injection}
\shortauthors{Huang et al.}

\begin{document}
\title{Low-frequency gravitational wave memory from gamma-ray burst afterglows with energy injection}

\correspondingauthor{Tong Liu}
\email{tongliu@xmu.edu.cn}

\author[0000-0002-4448-0849]{Bao-Quan Huang}
\affiliation{Department of Astronomy, Xiamen University, Xiamen, Fujian 361005, China}

\author[0000-0001-8678-6291]{Tong Liu}
\affiliation{Department of Astronomy, Xiamen University, Xiamen, Fujian 361005, China}

\author[0000-0002-0279-417X]{Li Xue}
\affiliation{Department of Astronomy, Xiamen University, Xiamen, Fujian 361005, China}

\author[0000-0002-1768-0773]{Yan-Qing Qi}
\affiliation{Department of Astronomy, Xiamen University, Xiamen, Fujian 361005, China}

\begin{abstract}
Ultrarelativistic gamma-ray burst (GRB) jets are strong gravitational wave (GW) sources with memory-type signals. The plateau (or shallow decay) phases driven by the energy injection might appear in the early X-ray afterglows of GRBs. In this paper, we investigate the GW signal as well as X-ray afterglow emission in the framework of GRB jets with energy injection, and both short- and long-duration GRBs are considered. We find that, regardless of the case, because of the antibeaming and time delay effects, a rising slope emerging in the waveform of GW signal due to the energy injection lags far behind the energy ejection, and the typical frequency of the characteristic amplitudes falls within a low-frequency region of $\sim10^{-4}-10^{-6} \,{\rm Hz}$; and we consider that the GW memory triggered by GRB jets with energy injection are previously unaware and the nearby GRBs with strong energy injection might disturb the measurement of the stochastic GW background. Such GW memory detection would provide a direct test for models of energy injection in the scenario of GRB jets.
\end{abstract}


\section{Introduction}

Gamma-ray bursts (GRBs) are generally classified as short-duration GRBs (SGRBs) corresponding to binary compact object mergers and long-duration GRBs (LGRBs) associated with the collapse of massive stars \citep[for reviews, see][]{Woosley2006,Nakar2007,Zhang2018book}. They are widely believed to be produced by ultrarelativistic jets launching from their central engines: a rapidly rotating black hole surrounded by a hyperaccretion disk \citep[see review by][]{Liu2017} or a massive millisecond magnetar \citep[e.g.,][]{Duncan1992,Usov1992,Dai1998,Zhang2001}. Commonly, following GRB prompt emission, there are long-lasting multiband afterglows that originate from the synchrotron radiation of electrons accelerated by forwarding shocks (FSs) generated due to the interaction between the jets and the circumburst material \citep[e.g.,][and references therein]{Zhang2006}.

With the enrichment of X-ray afterglow data collected by the Swift X-ray Telescope, a substantial fraction of GRBs is discovered for which a long plateau shape exists in the light curves of their early X-ray afterglows \citep[e.g.,][]{Zhang2006,Nousek2006,Yi2022}. The origin of this plateau, although widely investigated, is still highly debated because various models can explain the formation of the plateau to some extent, such as those models commonly involving energy injection \citep[e.g.,][]{Zhang2006,Nousek2006,Granot2006,Fan2013,Du2020,Hou2021}, two/multiple jet components \citep[e.g.,][]{Granot2006,Toma2006,Yamazaki2009}, a dominated reverse shock \citep{Uhm2007}, evolutive microphysical parameters \citep{Ioka2006}, slightly off-axis viewing angles \citep{Eichler2006,Beniamini2020}, phase transients of neutron stars \citep{Li2016,Hou2018}, a precessing jet \citep{Huang2021,Huang2022}, a wind environment \citep{Dereli2022}, and so on. In this case, other messengers seem to be necessary to further restrict models and to enhance our understanding of the physical origin of the plateau. Fortunately, the joint observation of the gravitational-wave (GW) events GW 170817 and GRB 170817A from the coalescence of two neutron stars has opened a new era of multimessenger astronomy \citep[e.g.,][]{Abbott2017,Goldstein2017}.

In the framework of GRBs, in addition to progenitors, i.e., the merger of binary compact objects, being GW sources, relativistic jets are also thought to be a strong GW source. Such sources with a memory-type GW signal have been widely studied. \cite{Segalis2001} applied a point mass approximation to investigate the GW signals from a relativistic GRB jet accelerating instantaneously and found that GW amplitudes, not like the electromagnetic emission beaming in the line-of-sight direction, are antibeamed (i.e., the antibeaming effect). \cite{Sago2004} studied GW signals produced by relativistic GRB jets with a finite half-opening angle based on the unified model in which the jets consisting of multiple subjects successively instantaneously accelerated; they found that the point mass approximation is not appropriate for jets with a half-opening angle larger than the inverse of the Lorentz factor, and many steps usually emerge in the waveform. \cite{Birnholtz2013} focused on GW signals from a single pulse of an accelerating GRB jet with a finite half-opening angle and with a finite acceleration timescale. In their work, both jet structures (uniform and structured) and the effect of equal arrival time surfaces are considered; it was found that under typical parameters of GRBs, the frequency and amplitudes of GW are $\lesssim 600 \,{\rm Hz}$ and $\sim 10^{-25}$ at a distance of 500 Mpc, respectively, and the peak amplitude for a uniform jet exhibits a smoother signal than the peak of a structured jet. \cite{Akiba2013} studied the effect of electromagnetic emission on the GW signal from a radiating and decelerating GRB jet and found that if the emission is partially anisotropic, there will be a change in the GW signals. \cite{Yu2020} calculated GW signals produced in the process of a relativistic GRB jet propagating into dense surroundings, a stellar envelope, or merger ejecta. \cite{Leiderschneider2021} computed the GW signals from accelerating GRB jets taking into account both time dimensions, i.e., the injection time and the acceleration time of the jet. \cite{Urrutia2022} carried out numerical simulations of relativistic jets penetrating through a massive, stripped-envelope star in the content of LGRBs and calculated the resulting GW signals.

If the plateaus in X-ray afterglows are caused by the additional energy from the central engine injected into the jets, GW from the jets would also be modulated and even possibly imprinted by the injection energy. If so, the models involving energy injection will be identified by their joint detections. Thus, in this work, we focus on investigating the GW signals from GRB jets with energy injection associated with the plateaus in the light curves of the X-ray afterglows. This paper is organized as follows. In Section 2, we introduce the model, including the dynamical description of GRB jets and the calculations of afterglows and GW emissions. The main results and corresponding toy model are presented in Section 3, and a brief summary is presented in Section 4.

\section{Model}
\subsection{Dynamics of GRB Jets}

In this work, we adopt a common scenario: namely, a relativistic GRB jet (i.e., an episode of ejecta, considered here) interacts with the circumburst medium, which results in production of an FS, and then the jet drives the FS propagating in the circumburst medium; meanwhile, continuous and long-lasting energy from the central engine injects into the jet. For this scenario, the dynamics of the jet can be described as follows \citep[e.g.,][]{Zhang2018book}:
\begin{widetext}
\begin{eqnarray}\label{eqn: Gamma}
\frac{d\Gamma}{dt_{\rm obs}} = -\frac{\Gamma(\Gamma^{2}-1)(\hat{\gamma}\Gamma-\hat{\gamma}+1)c^{2}\frac{dm}{dt_{\rm obs}}-\Gamma(\hat{\gamma}-1)(\hat{\gamma}\Gamma^{2}-\hat{\gamma}+1)(3U/R)\frac{dR}{dt_{\rm obs}}-(\Gamma^{2}-(\hat{\gamma}\Gamma^{2}-\hat{\gamma}+1)\xi)
\frac{dE_{\rm in}}{dt_{\rm obs}}}{\Gamma^{2}(M_{\rm ej}+m)c^{2}+(\hat{\gamma}^{2}\Gamma^{2}-\hat{\gamma}^{2}+3\hat{\gamma}-2)U},
\end{eqnarray}
\begin{eqnarray}\label{eqn: U}
\frac{dU}{dt_{\rm obs}}=(1-\epsilon)(\Gamma-1)c^{2}\dfrac{dm}{dt_{\rm obs}}-
(\hat{\gamma}-1)
\left( \frac{3}{R}\frac{dR}{dt_{\rm obs}}-\dfrac{1}{\Gamma}\frac{d\Gamma}{dt_{\rm obs}} \right)U+\frac{\xi}{\Gamma}\frac{dE_{\rm in}}{dt_{\rm obs}},
\end{eqnarray}
\begin{equation}\label{eqn: m}
\dfrac{dm}{dt_{\rm obs}}=\dfrac{c\beta}{1-\beta\cos\theta_{\rm ej}}4\pi R^{2}n m_{p},
\end{equation}
\begin{equation}\label{eqn: R}
\dfrac{dR}{dt_{\rm obs}}=\dfrac{c\beta}{1-\beta\cos\theta_{\rm ej}},
\end{equation}
\begin{equation}\label{eqn: E_in}
{\frac{dE_{\rm in}}{dt_{\rm obs}}}=
\left\{\begin{array}{ll}
P_{\rm in}(1-\beta)/(1-\beta\cos\theta_{\rm ej}),\quad 0 < T \leq T_{\rm end},\\ 0,\quad \quad \quad \quad \quad \quad \quad \quad \quad \quad \quad T_{\rm end} < T.
\end{array}
\right.
\end{equation}
\end{widetext}
where $\Gamma$, $U$, $m$, $R$, and $E_{\rm in}$ are the bulk Lorentz factor, internal energy, sweep-up mass from the circumburst medium by the FS, distance from the central source, and injection energy, respectively; and $\hat{\gamma}$, $\xi$, $\epsilon$, $n$, $m_{p}$, $P_{\rm in}$, $t_{\rm obs}$, and $T_{\rm end}$ represent the adiabatic index of the shocked medium \citep[e.g.,][]{Peer2012}, the fraction of the injection energy transforming into the internal energy, the overall fraction of the shock-dissipated energy that is transduced into radiation, the number density of the circumburst medium, the proton mass, the power of energy injection, the time measured in the observer frame, and the ending time of energy injection defined by an on-axis observer, respectively. $M_{\rm ej}=E_{\rm 0,iso}/\Gamma_{0}c^2$ is the initial jet mass, where $E_{\rm 0,iso}$ and $\Gamma_{0}$ are the initially isotropic energy and the initial Lorentz factor, respectively. In addition, $\beta=\sqrt{1-1/\Gamma^{2}}$ and $T=t_{\rm obs}-R(1-\cos\theta_{\rm ej})(1+z)/c$, where $c$ is the speed of light and $\theta_{\rm ej}$ is the angle between the line of sight and the movement direction of the jet or of the jet element. For the sake of simplicity, the circumburst medium is considered to be the interstellar medium (ISM), whose mass density remains constant at all times.

\subsection{Afterglow Radiation}

We assume a uniform jet with a half-opening angle $\theta_{\rm j}$ possessing a sector shape and neglect the evolution of $\theta_{\rm j}$ (i.e., the sideways expansion effect) because it is not important according to the results of numerical simulations \citep[e.g.,][]{Zhang2009}. To facilitate the description of the location of the moving jet, we set two Cartesian coordinate frames: (X,Y,Z) and (x,y,z), with the Z-axis being the line of sight, the z-axis corresponding to the symmetry axis of the jet, and the Y-axis coinciding with the y-axis \citep[e.g.,][]{Akiba2013}. The corresponding spherical coordinates of both frames are denoted by ($\theta$, $\phi$) and ($\vartheta$, $\varphi$), respectively. After defining the angle between the Z- and z-axes as the viewing angle $\theta_{\rm v}$, we can obtain the changing relation between these two spherical coordinates, which is expressed as
\begin{equation}\label{eqn: cos_phi}
\sin\theta\cos\phi=-\sin\theta_{\rm v}\cos\vartheta+\cos\theta_{\rm v}\sin\vartheta\cos\varphi,
\end{equation}
\begin{equation}\label{eqn: sin_phi}
\sin\theta\sin\phi=\sin\vartheta\sin\varphi,
\end{equation}
\begin{equation}\label{eqn: cos_theta}
\cos\theta=\cos\theta_{\rm v}\cos\vartheta+\sin\theta_{\rm v}\sin\vartheta\cos\varphi.
\end{equation}

We only consider the radiation mechanism responsible for the afterglows as the synchrotron radiation, whose power at the frequency $\nu'$ can be calculated by the following formula \citep{Rybicki1979}:
\begin{equation}\label{eqn: power}
P'_{\rm syn}({\nu}')=\frac{\sqrt{3} {\rm e}^3 B'}{m_{\rm e}c^2}\int\nolimits_{\gamma_{\rm e,min}'}^{\gamma_{\rm e,max}'}\bigg(\frac{dN_{\rm e}'}{d\gamma_{\rm e}'} \bigg) F \bigg(\frac{\nu'}{\nu_{\rm c}'} \bigg) d\gamma_{\rm e}',
\end{equation}
where $\rm e$ is the electron charge, $B'$ is the magnetic field, and $m_{\rm e}$ is the electron mass; and $\gamma'_{\rm e,min}$, $\gamma'_{\rm e,max}$, and $dN'_{\rm e} / d\gamma'_{\rm e}$ are the minimum Lorentz factor, the maximum Lorentz factor, and the energy spectrum of the shock-accelerated electrons, respectively \citep[e.g.,][]{Fan2008,Huang2021}. In addition, $F({\nu'}/{\nu_{\rm c}'})=(\nu'/\nu_{\rm c}') \int\nolimits_{\nu'/\nu_{\rm c}'}^{+ \infty} K_{5/3}(x) dx$, with $\nu_{\rm c}'=3 {\rm e} B' \gamma_{\rm e}'^{2}/4 \pi m_{\rm e} c$ being the critical frequency of electrons and Lorentz factors $\gamma'_{\rm e}$ and $K_{5/3}(x)$ being a modified Bessel function of order 5/3. Note that quantities with a prime are defined in the comoving frame of the jet.

To calculate the observed flux of the synchrotron radiation from the jet more accurately, we linearly divided the jet into $500\times1000$ point-like elements along the $\vartheta$- and $\varphi$-directions in the spherical coordinate system ($\vartheta$, $\varphi$). Therefore, the total observed flux can be obtained by summing the flux received from the point-like elements with the same observation time. Here, the total observed flux density can be written as \citep[e.g.,][]{Granot1999}
\begin{equation}\label{eqn: flux_density}
F_{\nu_{\rm obs}}=\frac{1+z}{4\pi D_{L}^2}{\int}\kern-4pt{\int}_{\kern-4pt(\rm EATS)}{P'_{\rm syn}(\nu'){D_d^3}d\Omega},
\end{equation}
where ``EATS'' is the equal-arrival time surface corresponding to the same observation time, $D_d = 1/\Gamma(1-\beta\cos\theta_{\rm ej})$ is the Doppler factor of the point-like elements, and $\nu^{\prime}=(1+z)\nu_{\rm obs}/D_d$, where $\nu_{\rm obs}$ is the observed frequency and $z$ is the redshift of the burst. $D_{L}$ represents the luminosity distance in the standard $\Lambda$CDM cosmology model with $\Omega_M=0.27$, $\Omega_\Lambda=0.73$, and $H_0=71~\rm km~s^{-1}~Mpc^{-1}$.

\begin{figure*}
\centering
\includegraphics[width=0.97\linewidth]{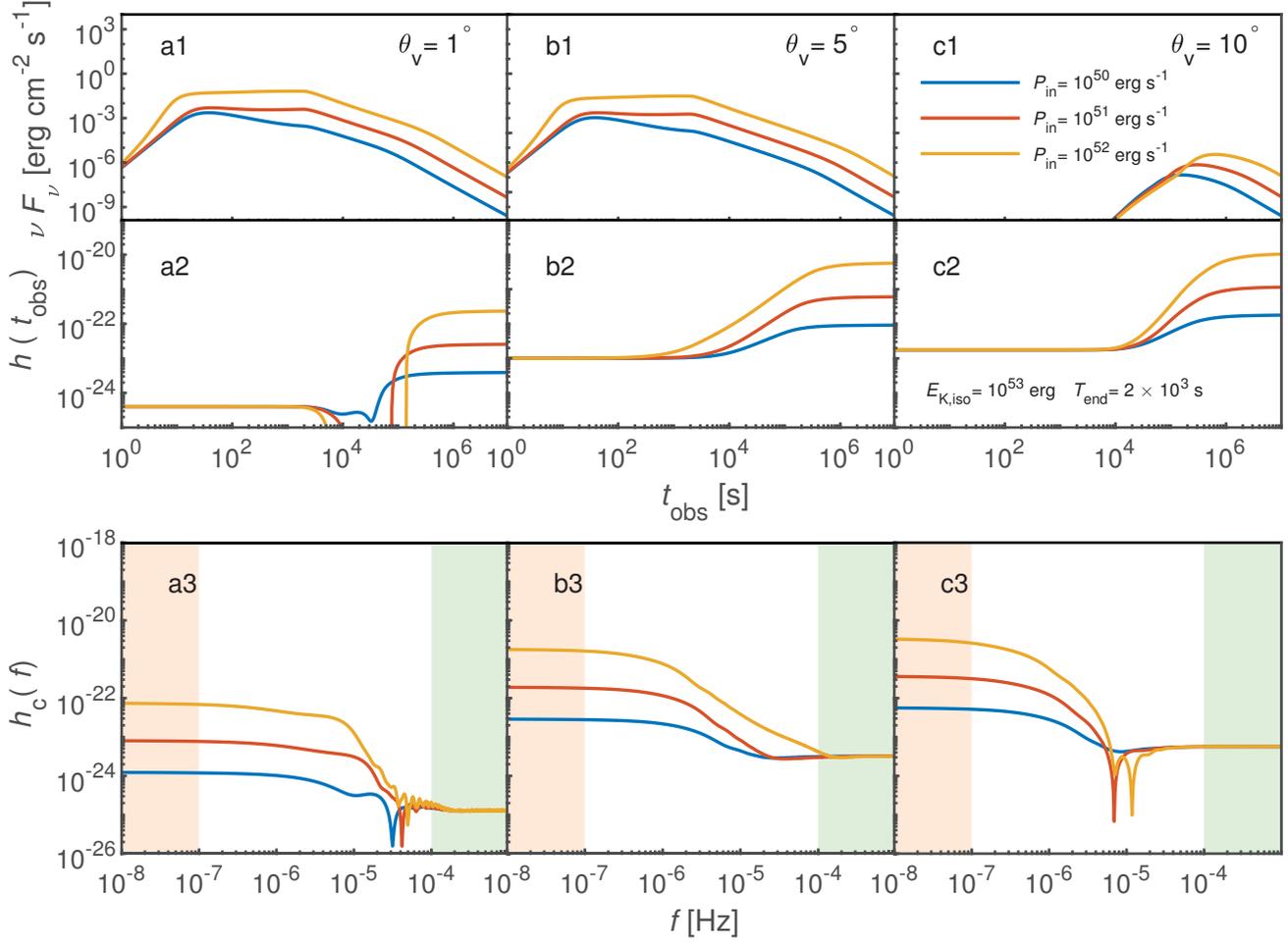}
\caption{GRB afterglow light curves, GW waveform, and characteristic amplitude $h_{\rm c}$ in Case I ($E_{\rm k,iso}=10^{53} \,{\rm erg}$ and $T_{\rm end}=2\times10^3 \,{\rm s}$) with the viewing angles $\theta_{\rm v}=1^{\circ}$, $5^{\circ}$, and $10^{\circ}$ corresponding to the first, second, and third column panels, respectively. The blue, red, and yellow lines denote the results for $P_{\rm in}=10^{50}$, $10^{51}$, and $10^{52} \,{\rm erg \,s^{-1}}$, respectively. In the bottom panels, the pink area denotes the GW sources, including stochastic background and supermassive binaries, and the green areas mainly contain massive binaries, resolvable and unresolvable galactic binaries, and extreme mass ratio inspirals \citep{Moore2015}.}
\end{figure*}

\subsection{GW signals}

We further compute the amplitudes, $h$, of the GW signals of the GRB jets with energy injection in the transverse-traceless (TT) gauge. The GW amplitudes generally include two polarization components: $h_{+}$ and $h_{\times}$. Based on the coordinate frame mentioned above, the $h_{\times}$ component vanishes due to axisymmetry, and hence, we simply denote $h=h_+$. For a jet element in a solid angle $\Delta \Omega$, the change of the amplitudes, $dh_1$, due to the shock-accelerated ISM, can be expressed as \citep[e.g.,][]{Segalis2001,Sago2004}
\begin{equation}\label{eqn: dh}
dh_1=\frac{2G}{c^4} \frac{dE}{D} \frac{\beta^2\sin^2\theta_{\rm ej}}{1-\beta \cos \theta_{\rm ej}}\cos 2 \phi_{\rm ej},
\end{equation}
where $G$ is the gravitational constant, $D$ is the distance between the jet element and the observer, and $(\theta_{\rm ej},\phi_{\rm ej})$ is the location of the element in the XYZ-frame. $dE=(c^2 \Gamma dm + \Gamma_{\rm eff} dU)\Delta \Omega /4\pi$ is the total increasing energy contained in the jet element, $\Gamma_{\rm eff} = (\hat{\gamma} \Gamma^2 - \hat{\gamma}+1) / \Gamma$, and $c^2 \Gamma dm$ and $\Gamma_{\rm eff} dU$ represent the obtained kinetic energy and internal energy of the newly shock-accelerated ISM, respectively.

\begin{figure*}
\centering
\includegraphics[width=0.46\linewidth]{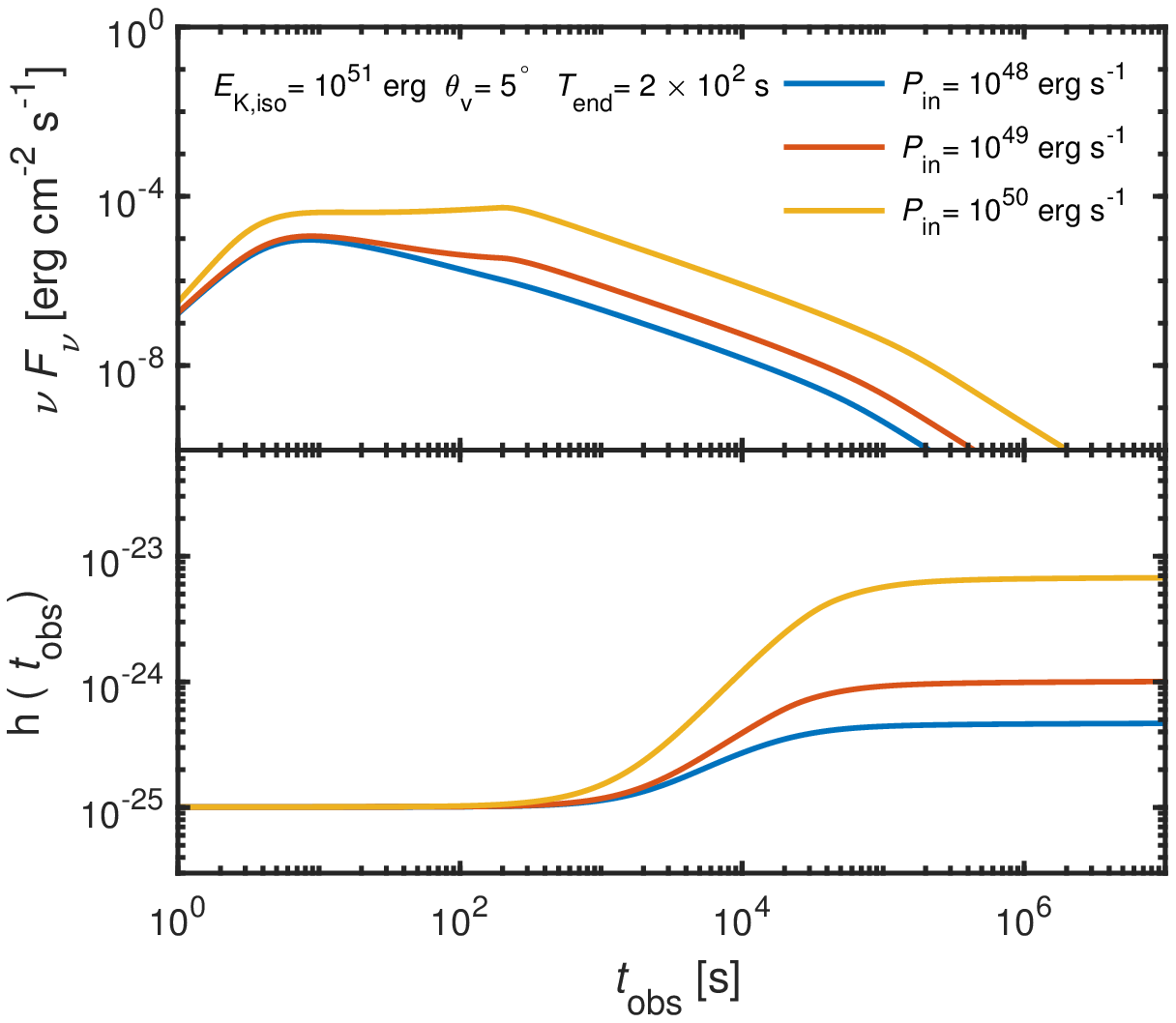}
\quad
\includegraphics[width=0.46\linewidth]{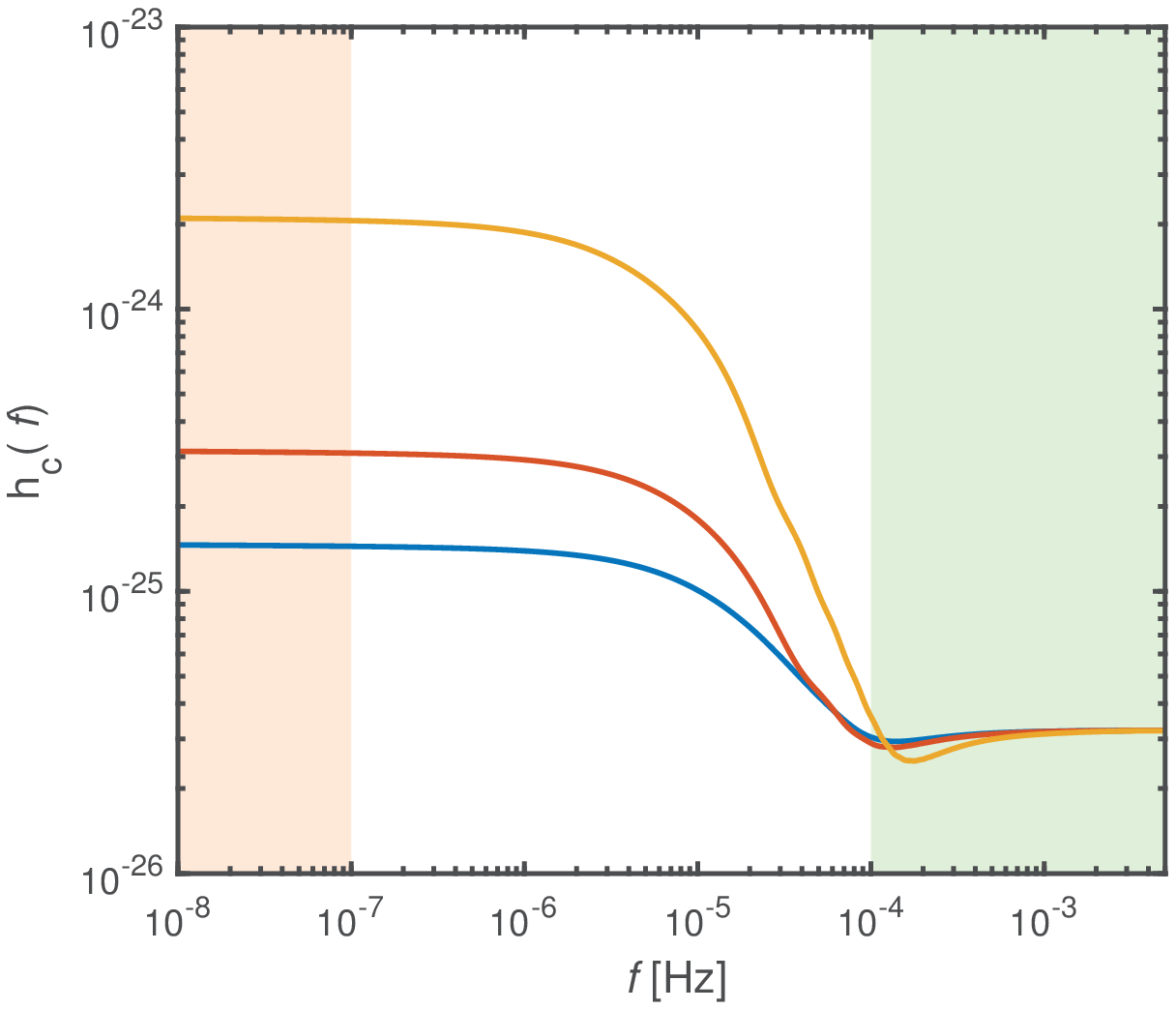}
\caption{Same as Figure 1 but for Case II ($E_{\rm k,iso}=10^{51} \,{\rm erg}$ and $T_{\rm end}=2\times10^2 \,{\rm s}$) with viewing angles $\theta_{\rm v}=5^{\circ}$. The blue, red, and yellow lines denote the results for $P_{\rm in}=10^{48}$, $10^{49}$, and $10^{50} \,{\rm erg \,s^{-1}}$, respectively.}
\end{figure*}

In the initial acceleration phase, the jets are assumed to be instantaneously accelerated \citep[e.g.,][]{Segalis2001,Sago2004}. Hence, according to Equation (\ref{eqn: dh}) and given total energy of a jet $E_{\rm j}$ to replace $E$, one can easily obtain the GW amplitudes $h_{\rm in}$ in this phase.

In the afterglow phase, the jet-driven FS continually sweeps up the ISM. These ISMs are accelerated (heated) and concentrated on the head of the jet, in which the shock-accelerated electrons produce synchrotron radiation. Simultaneously, the new energy from the central engine is injected into the jet from the back. In this scenario, the contribution of the radiated luminosity to the changes of the GW amplitudes can be described as \citep[e.g.,][]{Akiba2013}
\begin{equation}
\frac{d\dot{h}_2}{d \Omega}= \frac{G}{2\pi c^4} \frac{L_{\rm sy}}{D} \frac{\sin^2\theta_{\rm ej}}{1-\cos\theta_{\rm ej}} \cos 2 \phi_{\rm ej},
\end{equation}
where $``\dot{\quad}"$ denotes the time derivative, and $L_{\rm sy}$ is the luminosity of synchrotron radiation. The GW signals in GRB afterglows originate from the shock-accelerated ISM and the synchrotron photons, which is different to that produced in the initial acceleration phase caused by the acceleration of the jet material from zero to an ultrarelativistic velocity.

Then the change rate of the GW amplitudes in the observer frame can be calculated by accumulating the contributions arriving at the same time, which can be delineated as
\begin{equation}\label{equ: dh_dt}
\dot{h} (t_{\rm obs}) = {\int} \frac{d\dot{h}}{d\Omega} d\Omega.
\end{equation}
where $\dot{h}=\dot{h}_1+\dot{h}_2$.

Solving Equation (\ref{equ: dh_dt}) numerically, one can obtain the GW amplitudes $h(t_{\rm obs})$. In addition, the characteristic amplitudes evolving in the frequency domain are expressed as \citep[e.g.,][]{Sago2004}
\begin{equation}\label{equ: hc}
h_{\rm c}(f)=2f|\widetilde{h}_{\rm in}(f)+\widetilde{h}_{\rm af}(f)|,
\end{equation}
where
\begin{equation}\label{equ: hf_in}
\widetilde{h}_{\rm in}(f)= \int_{-\infty}^{\infty} h_{\rm in} e^{-2\pi i f t_{\rm obs}} dt_{\rm obs},
\end{equation}
and
\begin{equation}\label{equ: hf_af}
2\pi i f \widetilde{h}_{\rm af}(f)= \int_{-\infty}^{\infty} \dot{h}(t_{\rm obs}) e^{-2\pi i f t_{\rm obs}} dt_{\rm obs}.
\end{equation}

\section{Results}

We calculate the light curves of X-ray afterglows in the observed energy bands from 0.3 to 10 ${\rm keV}$ and GW amplitudes from a GRB jet with energy injection. Two cases are considered, i.e., LGRBs (Case I) and SGRBs (Case II). According to the results from sample statistics \citep[e.g.,][]{Lu2014}, we set $E_{\rm k,iso}=10^{53} \,{\rm erg}$, $T_{\rm end}=2\times 10^3 \,{\rm s}$, $P_{\rm in}=10^{50},10^{51}$, and $10^{52} \,{\rm erg \,s^{-1}}$ for LGRBs, and $E_{\rm k,iso}=10^{51} \,{\rm erg}$, $T_{\rm end}=2\times 10^2 \,{\rm s}$, $P_{\rm in}=10^{48},10^{49}$, and $10^{50} \,{\rm erg \,s^{-1}}$ for SGRBs. Here, $E_{\rm k,iso}$ is the isotropic kinetic energy, which is assumed to be numerically equal to the isotropic energy, i.e., $E_{\rm 0,iso}=E_{\rm k,iso}$, since, generally, the jet energy is dominated by kinetic energy in the initial phase of the afterglows for a fireball scenario. Moreover, considering the loss in energy caused by the prompt emission being significantly small, we set $E_{\rm j} \approx E_{\rm 0,iso}$. Except for the above mentioned parameters, the other basic parameters are set to be the same value for both cases. The effect of the injection energy is considered to primarily boost the kinetic energy of the jet, i.e., $\xi=0$. The radiation efficiency is defined as $\epsilon=\epsilon_{\rm rad}\epsilon_{\rm e}$ \citep[e.g.,][]{Nava2013}; $\epsilon_{\rm e}$ is the fraction of the internal energy that is shared by electrons, which is set as 0.1; $\epsilon_{\rm rad}$ is the fraction of the electron energy that is radiated, which is expressed as $\epsilon_{\rm rad}=(\gamma_{\rm m}/\gamma_{\rm c})^{p-2}$, where $\gamma_{\rm m}$ is the minimum electron Lorentz factor, $\gamma_{\rm c}$ is the critical electron Lorentz factor, and $p=2.5$ is the power-law index of the electron energy spectrum. In addition, $\Gamma_0=200$, $\theta_{\rm j}=5^{\circ}$, $n=1.0$, and $D\approx D_{L}=1 \,{\rm Mpc}$ are adopted.

\subsection{Case I}

\begin{figure*}
\centering
\includegraphics[width=0.45\linewidth]{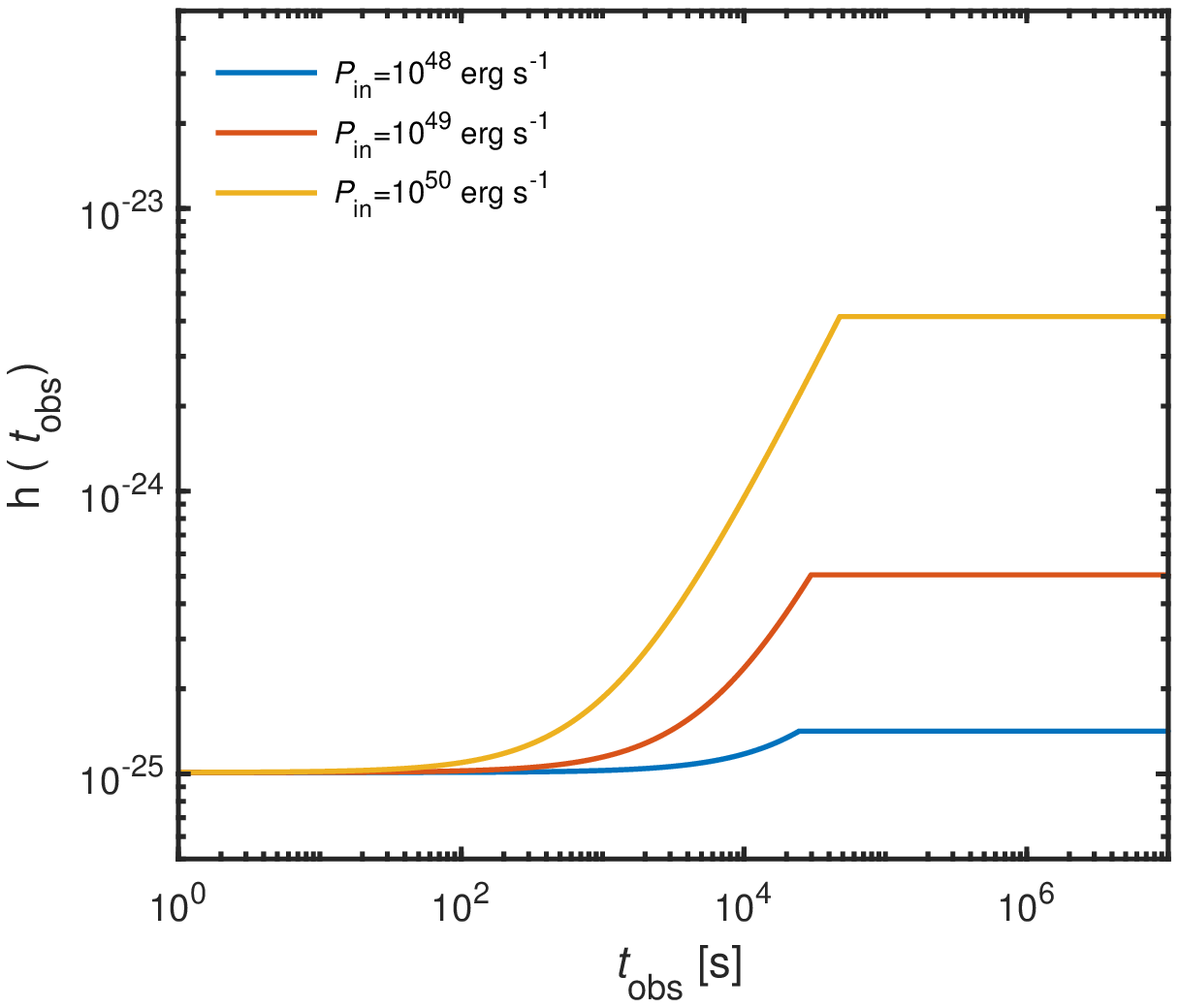}
\includegraphics[width=0.45\linewidth]{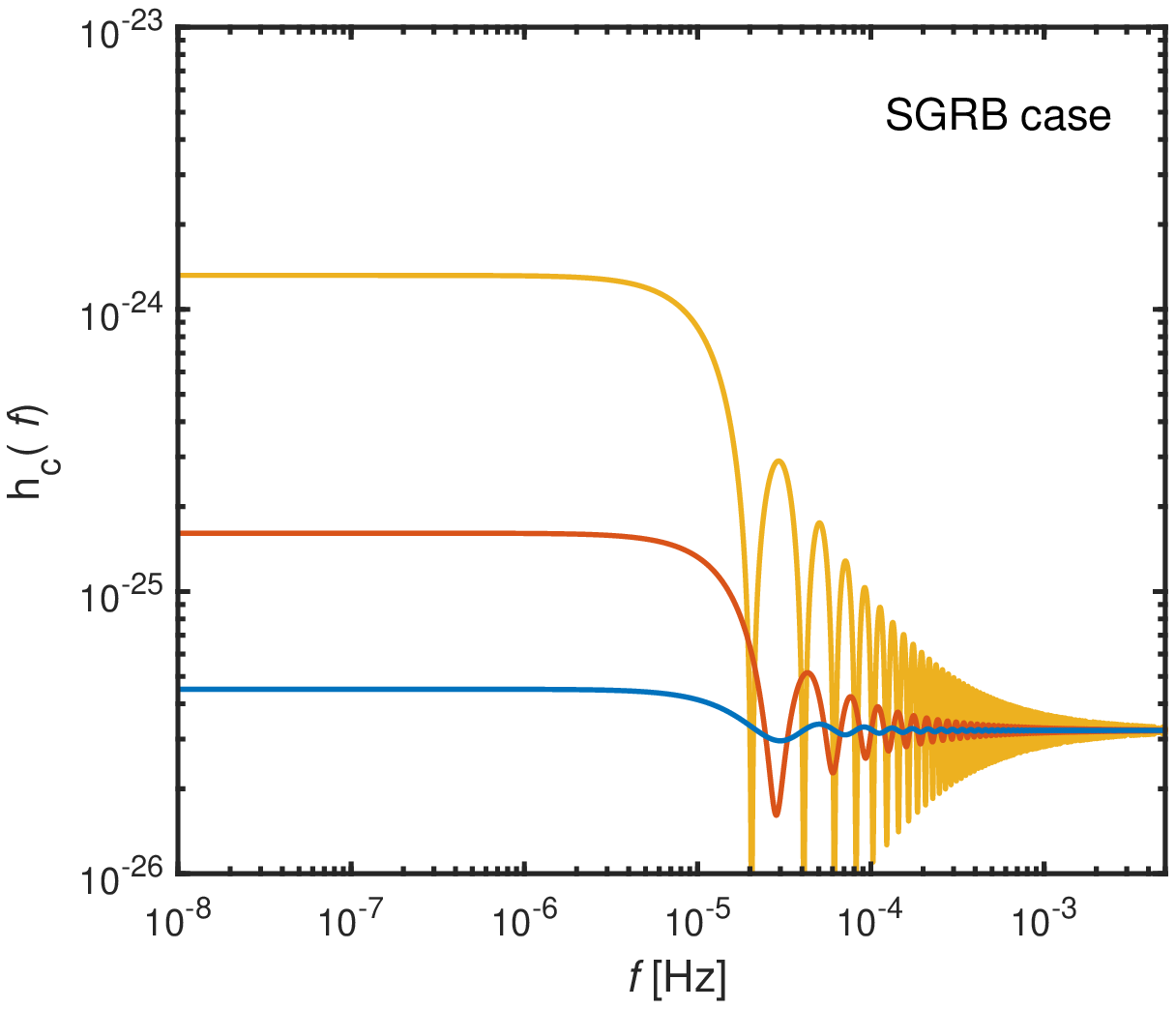}
\caption{GW waveforms and characteristic amplitudes for SGRB cases in toy model.}
\end{figure*}

Figure 1 shows the results for Case I (LGRBs). Panels with serial numbers a, b, and c denote the results for viewing angles $\theta_{\rm v}= 1^{\rm \circ}, 5^{\circ}$ and $10^{\circ}$, respectively. Panels with serial numbers 1, 2, and 3 represent the light curves of the X-ray afterglows, the GW amplitudes in the time domain, and the characteristic amplitudes in the frequency domain, respectively. The blue, red, and yellow lines correspond to the results for $P_{\rm in}=10^{50}$, $10^{51}$, and $10^{52} \,{\rm erg \, s^{-1}}$, respectively.

An expected plateau shape appears in the light curves of the X-ray afterglows due to the energy injection, as shown in Panels (a1) and (b1). However, one can find that the appearance of the plateaus depends on the power of the energy injection $P_{\rm in}$. When the power reduces, e.g., $P_{\rm in}= 10^{50} \,{\rm erg \, s^{-1}}$, the plateaus disappear, and a shallow decay shape appears. Since a uniform jet with a sharp edge is considered, the behaviors of the light curves are similar for $\theta_{\rm v}\leq \theta_{\rm j}$, and the light curves present a bump structure for $\theta_{\rm v} > \theta_{\rm j}$, as exhibited in Panel (c1). In addition, the values of the fluxes are significantly high because of the close distance, $D_{L}=1 \,{\rm Mpc}$.

The amplitudes of GW signal originated from the initial acceleration phase exhibit no obvious changes before $\sim10^4 \,{\rm s}$ since the increased energy is far less than the initial kinetic energy for the jet in the early stage of the afterglows and the off-axis observation causes a time delay. After that, a visible growth emerges in the amplitudes owing to the energy injection. The growths of the amplitudes become increasingly significant as the power of the injection energy increases. They reach maximum values in a time range of $10^4 \sim 10^6 \,{\rm s}$ and remain essentially constant after the maximum value. As mentioned above, the jet energy in initial phase, $E_{\rm j}$, is set to be equal to $E_{\rm k,iso}$ in value, and the jet energy in afterglow phase is $E_{\rm k,iso}+P_{\rm in} \times T_{\rm end}$. It is obvious that the jet energy in afterglow phase is larger than that in initial phase. Thus the appearance of rising slopes in the curves of the amplitudes are reasonable. In addition, there exists a negative $h(t_{\rm obs})$ when $\theta_{\rm v}=1^{\circ}$, as shown in Panel (a2). This feature, as also shown in Figure (7) of \cite{Birnholtz2013}, is attributed to the negative contribution from the region of the jet around the observer direction. Due to the antibeaming effect, the GW amplitudes in $\theta_{\rm v}=1^{\circ}$ are significantly smaller than those in $\theta_{\rm v}=5^{\circ}$ and $\theta_{\rm v}=10^{\circ}$, but not approximately zero. Moreover, the combination of the antibeaming effect \citep{Segalis2001} and the time delay effect \citep{Birnholtz2013} leads to the plateaus in the afterglow light curves and the rising slope of the GW amplitudes not being produced in the same range of time for $\theta_{\rm v}\leq \theta_{\rm j}$; however, a similar rise shape can appear in both for $\theta_{\rm v} > \theta_{\rm j}$, as shown in Panels (c1) and (c2).

In the bottom panels, pink and green areas denote the characteristic amplitudes of GW sources with very low frequency and low frequency, respectively; the former includes stochastic background and supermassive binaries; the latter mainly involves massive binaries, resolvable and unresolvable galactic binaries, and extreme mass ratio inspirals \citep[][or the GW Sensitivity Curve Plotter\footnote{http://gwplotter.com/}]{Moore2015}. One can find that two plateaus appear in the curves of the characteristic amplitudes. One with a smaller value in higher frequency corresponds to the initial acceleration phase and the other with a larger value in lower frequency mainly originates from the afterglow phase. Moreover, there is a continuous increase in the characteristic amplitudes between both plateaus from right to left in frequency, and the corresponding frequencies are within a low-frequency range of $10^{-4} - 10^{-6}$ Hz, which is a blank area between low frequency and very low frequency for known GW sources \citep[e.g.,][]{Cutler2002}. Interestingly, the very low frequency GW memory from very nearby GRB jets with very strong energy injection could influence the measurement of the stochastic background, although they are very rare events. In addition, regardless of the change in viewing angle, the characteristic amplitude increases with increasing injection energy.

\begin{figure}
\centering
\includegraphics[width=0.95\linewidth]{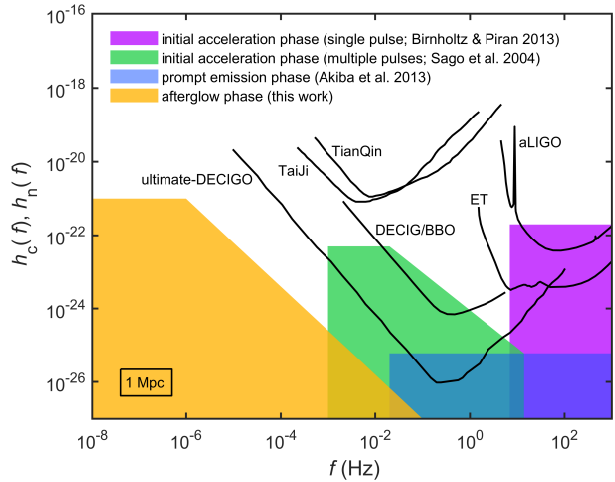}
\caption{The characteristic amplitude scopes of GW signals from GRB jets in different stages with a distance of 1 Mpc. The purple, green, blue, and yellow shaded areas represent the initial acceleration phase (single pulse), the initial acceleration phase (multiple pulses), the prompt emission phase, and the afterglow phase, respectively. The black lines are the sensitivity lines of detectors including aLIGO, ET, TianQin, Taiji, DECIGO/BBO, and ultimate-DECIGO.}
\end{figure}

\subsection{Case II}

In Case II (SGRBs), we only consider the situation of $\theta_{\rm v}=5^{\circ}$, and the corresponding results are shown in Figure 2. Comparing the results of $\theta_{\rm v}=5^{\circ}$ in Case I, one can observe that similar plateau (or shallow decay) shapes appear in the light curves, the rising shapes emerge in the curve of the GW amplitudes, and the characteristic amplitudes gradually reach maximum values in the frequency region of $\sim 10^{-4}-10^{-6} \, {\rm Hz}$; however, there are still some differences. For instance, the duration of the plateaus (or shallow decays) is shorter, the maximum-value of the GW amplitudes occurs earlier, and the frequency of the characteristic amplitudes reaching maximum-value is higher due to the shorter injection timescale in Case II. In addition, for $P_{\rm in}= 10^{48} \, {\rm erg\, s^{-1}}$, the effect of the injection energy on the light curves is hardly significant, but a rising shape still can be seen in the corresponding curve of GW amplitudes.

\subsection{Toy model}

The zero-frequency limit (ZFL) of the energy spectrum of gravitational radiation, $dE/df d\Omega \mid_{f=0}$, reveals the existence of the GW memory \citep[e.g.,][]{Smarr1977}. In the scenario of GRBs, one can find $dE/df d\Omega |_{f = 0} \propto |h_{\infty}|^2$, where $h_{\infty}$ denotes the GW amplitude of GRB jets at $t \rightarrow + \infty$. Here, according to Equation (\ref{eqn: dh}) and adopting a point approximation for the jet, we define $h_{\infty} = 2 G E_{\rm to} \beta^2 \sin^2 \theta_{\rm ej} / c^4 D (1 - \beta \cos \theta_{\rm ej})$, where $\beta \sim 1$, $\theta_{\rm ej} = 2 \theta_{\rm v}$, and $E_{\rm to} = (E_{\rm j} + E_{\rm in}) (1 - \cos \theta_{\rm j}) / 2$ is the total increasing energy for the jet in whole evolutive process. Although the moving jet eventually stops due to the increasing mass of shocked ISM, the amplitude does not change so much. This is because the GW memory does not rely on the form of energy and the total energy almost all transforms into photon emission whose memory remains when $t \rightarrow + \infty$. Hence, the definition of $h_{\infty}$ should be reasonable. Considering the SGRB case with $P_{\rm in} = 10^{50} \,{\rm erg \, s^{-1}}$, we can obtain $h_{\infty} \simeq 4.3 \times 10^{-24}$. This result further verifies our numerical results with the same parameters shown in Figure 2.

Here we can construct an analytical waveform as follows:
\begin{equation}\label{eqn: h_analytical}
{h_{\rm a}(t_{\rm obs})}=
\left\{\begin{array}{ll}
0,\quad \quad \quad \quad \quad \quad \quad \quad \quad t_{\rm obs} \leq 0,\\
h_{\rm in} + h_{\rm m} \left(t_{\rm obs}/t_{\rm m} \right),\quad 0 < t_{\rm obs} \leq t_{\rm m},\\
h_{\rm in} + h_{\rm m}, \quad \quad \quad \quad \quad \quad t_{\rm m} < t_{\rm obs},
\end{array}
\right.
\end{equation}
where $t_{\rm m} = T_{\rm end} + R_{\rm end}(1-\cos \theta_{\rm ej})(1+z) / c$ with $R_{\rm end}$ being the distance of the jet to the source at $T_{\rm end}$ and $h_{\rm m} = G P_{\rm in} T_{\rm end} \beta^2 \sin^2 \theta_{\rm ej} (1-\cos \theta_{\rm j}) / c^4 D (1 - \beta \cos \theta_{\rm ej})$ with a order of $ \sim 5 \times 10^{-24} (\frac{P_{\rm in}}{10^{50} \, {\rm erg \, s^{-1}}}) (\frac{T_{\rm end}}{200 \, {\rm s}}) (\frac{D}{1 \, {\rm Mpc}})^{-1}$. Then, we calculate the Fourier component of this waveform and obtain
\begin{equation}\label{eqn: hf2_analytical}
|\widetilde {h}_{\rm a}(f)|^2=4 a^2 \sin^4 (\pi f t_{\rm m}) + [a \sin (2 \pi f t_{\rm m}) + b]^2,
\end{equation}
where $a = h_{\rm m} / 4 \pi^2 f^2 t_{\rm m}$ and $b = h_{\rm in}/2 \pi f$. Applying Equations (\ref{eqn: h_analytical}) and (\ref{eqn: hf2_analytical}), we further compute the GW waveform and characteristic amplitudes of the GW signal in SGRB case and the corresponding results are shown in Figure 3, which are consistent with our numerical results.

\section{Summary}

The observational X-ray plateaus arising in the light curves of the X-ray afterglows might imply long-lasting energy from the central engine injected into the jets. In this work, we investigate the GW signal from GRB jets with energy injection, as well as the light curves of X-ray afterglows. As a result, a rising slope emerging in the waveform of GW signal due to the energy injection lags far behind the energy ejection. The typical GW frequency is within a low-frequency region. The detection of these unique GW memory would provide a direct test for models of energy injection.

Moreover, we summarize the characteristic amplitude scopes of GW signals from GRB jets in different stages at a distance of 1 ${\rm Mpc}$, as shown in Figure 4. The yellow area denotes the result calculated in this work, and other colors are representative results corrected to 1 ${\rm Mpc}$ in distance from previous literature \citep[e.g.,][]{Sago2004,Birnholtz2013,Akiba2013,Wei2020}. The black lines represent the sensitivity lines (the noise amplitudes $h_{\rm n}$) of some detectors, including the advanced Laser Interferometer Gravitational-Wave Observatory (aLIGO), Cosmic Explorer (CE), Taiji Program in Space, TianQin Project, Decihertz Interferometer Gravitational Wave Observatory/Big Bang Observer (DECIGO/BBO), and ultimate-DECIGO \citep[e.g.,][or the GW Sensitivity Curve Plotter]{Moore2015}. One can determine that the GW signals from GRB jets in the initial acceleration phase as well as the prompt emission phase are expected to have characteristic frequencies of $\sim10^{-2} \, - \, 10^{3} \,{\rm Hz}$ and a characteristic amplitude at a level higher than, or close to, the sensitivity of the detectors. Unfortunately, the GW signals in the afterglow phase cannot be detected by these detectors because of the lower frequency, even if the characteristic amplitude is slightly higher than those in the initial acceleration/prompt phase. Regardless, the GRB jets with energy injection are distinctive new sources whose GWs are previously unaware in a blank GW band. Although the nearby GRBs with strong energy injection are the very rare events, their GW signal might disturb the measurement of the stochastic GW background.

\acknowledgments
We thank the anonymous referee for very useful comments and suggestions. This work was supported by the National Natural Science Foundation of China under grants 12173031 and 12221003.

\end{document}